\documentclass[pra,superscriptaddress,twocolumn,amsmath,amssymb]{revtex4}

\usepackage{natbib}

\usepackage{graphicx}

\usepackage{amssymb}

\begin{document}

\title{The effects of a pyrrolidine functional group on the magnetic properties of ${\rm N@C_{60}}$}

\author{Jinying Zhang}
\affiliation{Department of Materials, Oxford University, Oxford
OX1 3PH, United Kingdom}

\author{John~J.~L.~Morton}
\affiliation{Department of Materials, Oxford University, Oxford
OX1 3PH, United Kingdom}
\affiliation{Clarendon Laboratory, Department of Physics, Oxford
University, Oxford OX1 3PU, United Kingdom}

\author{Mark~R.~Sambrook}
\affiliation{Department of Materials, Oxford University, Oxford
OX1 3PH, United Kingdom}

\author{Kyriakos~Porfyrakis}
\email{kyriakos.porfyrakis@materials.ox.ac.uk}
\affiliation{Department of Materials, Oxford University, Oxford
OX1 3PH, United Kingdom}

\author{Arzhang~Ardavan}
\affiliation{Clarendon Laboratory, Department of Physics, Oxford
University, Oxford OX1 3PU, United Kingdom}

\author{G.~Andrew~D.~Briggs}
\affiliation{Department of Materials, Oxford University, Oxford
OX1 3PH, United Kingdom}

\begin{abstract}
A new stable pyrrolidine functionalized fullerene derivative, ${\rm
C_{69}H_{10}N_2O_2}$, has been synthesized, purified by high
performance liquid chromatography, and characterized by MALDI mass
spectrometry, ultraviolet-visible spectroscopy, Fourier transform
infrared, $^1$H and $^{13}$C nuclear magnetic resonance. The
magnetic properties of the analogous endohedral species have been
studied by both continuous wave (CW) and pulsed EPR. CW-EPR spectra
indicated an anisotropic hyperfine interaction and a permanent
zero-field-splitting (ZFS). Both CW and pulsed EPR showed the ZFS
parameter $D_{eff}$ to be around 17~MHz. Pulsed EPR revealed a
biexponential decay in both T$_1$ and T$_2$, yielding a molecular
tumbling correlation time ${\tau_c}$ of $31.4\pm2.5$~ps.
\end{abstract}


\maketitle

\section{Introduction}
\label{} Fullerenes which encapsulate atomic nitrogen (such as ${\rm
N@C_{60}}$) have been shown to contain an electron spin benefiting
from long relaxation times and narrow linewidths in electron
paramagnetic resonance (EPR)~\cite{murphy96}. Such properties make
${\rm N@C_{60}}$ an attractive candidate building block for a
quantum information processing device~\cite{harneit02} and a
powerful spin probe for measuring environments such as the interior
of nanotubes~\cite{dinseNT}.

Relaxation in unmodified ${\rm N@C_{60}}$ has been found to be
driven by a vibrational Orbach mechanism~\cite{john06}. If the
fullerene is modified by a functional group in such a way as to
introduce a zero-field-splitting (ZFS), the fluctuation of this ZFS
due to molecular tumbling in liquid solution provides an additional
relaxation pathway characterized by the magnitude of the ZFS, and
the rotational correlation time of the molecule. Thus, by measuring
T$_1$ and T$_2$, respectively the spin-lattice and phase memory
relaxation times, it is possible to probe the molecular dynamics of
fullerene derivatives in solution.

The fluctuating ZFS mechanism is expected to produce a biexponential
decay in both T$_1$ and T$_2$, assuming the molecule is not in the
`fast-tumbling' limit. Such a biexponential decay was not observed
in earlier studies on fullerene derivatives, reportedly due to
limited signal to noise and a relatively fast rotational correlation
time. The effect of nuclear spins in the solvent may also have
complicated the analysis by adding additional relaxation
pathways~\cite{relaxtol}. The observation of a biexponential decay
is simplified by choosing a bulky functional group, such as
pyrrolidine, to slow down the rotation further.

In this letter, we describe the synthesis of a new
pyrrolidine-functionalized fullerene derivative, ${\rm
C_{69}H_{10}N_2O_2}$, and its analogous endohedral species, ${\rm
N@C_{69}H_{10}N_2O_2}$. The product is purified by high performance
liquid chromatography (HPLC), characterized by MALDI mass
spectrometry, ultraviolet-visible (UV-vis), Fourier transform
infrared (FTIR), $^1$H and $^{13}$C nuclear magnetic resonance (NMR)
spectroscopies. We measure the ZFS and hyperfine interaction of the
electron spin arising from the addition of the functional group to
${\rm N@C_{60}}$ using continuous wave (CW) EPR. In contrast to
previous reports of fullerene derivatives~\cite{franco06,dietel99},
including pyrrolidine-functionalized fullerenes, we observe an
anisotropy in the $^{14}$N hyperfine interaction in this species.
Finally, we use pulsed EPR to measure T$_1$ and T$_2$ and thereby
deduce the ZFS $D_{eff}$ and rotational correlation time $\tau_c$ of
${\rm N@C_{69}H_{10}N_2O_2}$ in carbon disulfide solution at room
temperature.

\section{Experimental Details}
\textbf{${C_{69}H_{10}N_2O_2}$}: A mixture of ${\rm C_{60}}$ (MER
corporation, 99.5+$\%$) (70.0~mg, 0.1~mmol), 4-nitrobenzaldehyde
(98$\%$, Aldrich) (75.5~mg, 0.5~mmol), and N-methylglycine (98$\%$,
Aldrich) (17.8~mg, 0.2~mmol) in toluene (50~mL) was heated under
reflux for 2~h under a nitrogen  atmosphere. The resulting reaction
mixture was collected, filtered, and purified by HPLC (5PBB,
20$\times$250~mm, toluene eluent, 18 mL/min) to give the pure
product in 31$\%$ yield. MALDI \emph{m/z}: 898.2 (M$^+$, ${\rm
C_{69}H_{10}N_2O_2}$ requires 898.1). UV-vis (toluene):
$\lambda_{max}$~nm 314 (${\rm \varepsilon=3658.9~m^2}$mol$^{-1}$),
433 (${\rm \varepsilon=300.66~m^2}$mol$^{-1}$). FTIR (KBr):
527~cm$^{-1}$ (s, ${\rm C_{60}}$ cage); 790-859~cm$^{-1}$ (w,
out-of-plane aromatic C-H bending); 1343~cm$^{-1}$ (s, nitro group
symmetric stretching vibration), 1424~cm$^{-1}$, 1462~cm$^{-1}$, and
1600~cm$^{-1}$ (w, aromatic C-C stretching); 1522~cm$^{-1}$ (s,
nitro group asymmetric stretching vibration); 2776-2941~cm$^{-1}$
(w, br, aromatic C-H stretching)(s-strong, w-weak, br-broad).
$^1$H-NMR(500~MHz, ${\rm CS_2}$:${\rm CDCl_3}$=3:1):
$\delta$(ppm)=2.83 (s, 3H, H-3), 4.33 (d, J=9.5Hz, 1H, H-2), 5.02
(d, J=9.5Hz, 1H, H-2), 5.07 (s, 1H, H-1), 7.10 (d, J=8Hz, 1H, H-5 or
H-9), 7.18 (d, J=7Hz, 1H, H-9 or H-5), 8.04 (band, 1H, H-6 or H-8),
8.30 (d, J=9Hz, 1H, H-8 or H-6). $^{13}$C NMR(125.8~MHz, ${\rm
CS_2}$:${\rm CDCl_3}$=3:1): $\delta$(ppm)=155.65, 153.48, 152.37,
151.85, 148.06, 147.45, 147.42, 146.46, 146.44, 146.36, 146.30 (1C,
C-7), 146.27, 146.22, 146.10, 145.79, 145.75, 145.55, 145.53,
145.50, 145.48, 145.47, 145.39, 145.35, 145.33, 144.84, 144.65,
144.55, 144.42, 143.32, 143.20, 142.89, 142.80, 142.78, 142.73,
142.34, 142.31, 142.30, 142.27, 142.18, 142.07, 141.98, 141.90,
141.85, 141.73, 140.45, 140.19, 139.75, 137.26, 136.42, 136.23,
135.68, (50 peaks from \emph{sp$^2$} $C_{60}$ carbon) 130.10 (2C,
C-5 and C-9), 128.40 (1C, C-4), 124.02 (2C, C-6 and C-8), 82.75 (1C,
C-1), 70.18 (1C, C-2), 69.06 (1C, C-11), 68.05 (1C, C-10), 40.14
(1C, C-3).

\textbf{${N@C_{69}H_{10}N_2O_2}$}: Method as for ${\rm
C_{69}H_{10}N_2O_2}$ using ${\rm N@C_{60}/C_{60}}$ (1/100) as
starting material, the reaction was carried out in the absence of
light. The nitrogen EPR signal intensity of the product mixture is
approximately 73$\%$ of the initial signal.

EPR sample preparation:  N@${\rm C_{69}H_{10}N_2O_2}$/${\rm
C_{69}H_{10}N_2O_2}$ (around 7/1000) sample was dissolved in ${\rm
CS_2}$, degassed, and sealed in quartz EPR tubes (3~mm diameter).

CW EPR measurements were performed on a Magnettech Miniscope MS200
using a 2~G modulation, 270~s scan time, and 99 scans at room
temperature. Spectral simulation was performed using the EasySpin
software package~\cite{easyspin}.

Pulsed EPR measurements were performed using an X-band Brucker
Elexsys580e spectrometer. Spin lattice relaxation time T$_1$ and
spin dephasing time T$_2$ were measured by an inversion recovery
sequence and Hahn echo sequence, respectively. The $\pi/2$ and $\pi$
pulse durations were 56 and 112~ns respectively. Phase cycling was
used to eliminate the contribution of unwanted free induction decay
(FID) signals.

\section{Results and Discussion}
\subsection{Synthesis and Characterization}
\begin{figure}[t] \centerline
{\includegraphics[width=3.2in]{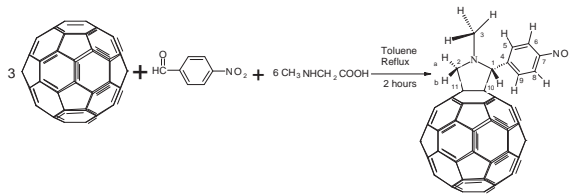}} \caption{Scheme of the synthesis of the pyrrolidine
functionalized fullerene derivative.}
\label{scheme}
\end{figure}

\begin{figure}[t] \centerline
{\includegraphics[width=3.2in]{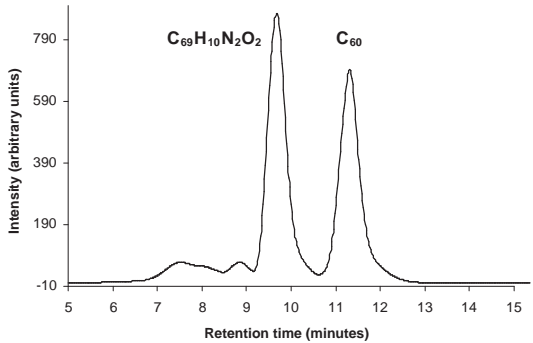}} \caption{HPLC chromatogram of the product mixture from
synthesis, HPLC (5PBB, 20$\times$250~mm, toluene eluent, 18 mL/min)}
\label{hplc}
\end{figure}

A nitrobenzene functionalized fullerene, ${\rm C_{69}H_{10}N_2O_2}$,
was synthesized according to the pyrrolidine functionalization
method~\cite{maggini93}~(Fig.\ref{scheme}). Briefly,
\emph{p}-nitrobenzaldehyde, N-methylglycine, and ${\rm C_{60}}$ were
heated under reflux in toluene for 2~h and subsequently purified by
HPLC (Fig.\ref{hplc}). ${\rm C_{69}H_{10}N_2O_2}$ elutes
approximately 2 minutes earlier than $\rm C_{60}$. Recycling mode
HPLC was used to purify the product to greater than 95$\%$.

\begin{figure}[t] \centerline
{\includegraphics[width=3.2in]{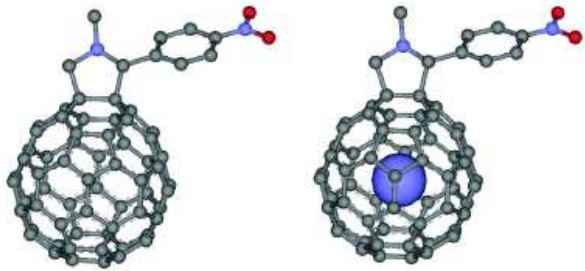}} \caption{Structural models of fullerene derivatives: ${\rm
C_{69}H_{10}N_2O_2}$ and ${\rm N@C_{69}H_{10}N_2O_2}$.}
\label{model}
\end{figure}

MALDI mass spectrometry revealed the expected molecular ion peak at
\emph{m/z} 898.2 and an isotopic distribution pattern in agreement
with that calculated. UV-vis spectrum features at 314~nm and 433~nm
are indicative of pyrrolidine functionalization and consistent with
other pyrrolidinization
compounds~\cite{maggini93,rodriguez04,maggini95,ros96}. Nitro group
symmetric and asymmetric stretching vibrations were observed in FTIR
at 1343~cm$^{-1}$ and 1522~cm$^{-1}$. Finally, $^1$H and $^{13}$C
NMR provide confirmation of the structure. Four different proton
environments were found in the pyrrolidine ring. The two proton
environments, H-2a and H-2b, were found to be stereo different and
consistent with other pyrrolidinization
compounds~\cite{maggini95,ros96}. Four inequivalent aromatic proton
environments were also observed, one of which was found to be broad
resonance consistent with previously reported
compound~\cite{rodriguez04}. The $^{13}$C NMR spectrum of compound
revealed 50 \emph{sp$^2$} and 2 \emph{sp$^3$} environments relating
to the ${\rm C_{60}}$ cage and an additional 3 pyrolidine
\emph{sp$^3$} and 4 nitrobenzene \emph{sp$^2$} carbon environments
as expected.

The endohedral fullerene species ${\rm N@C_{69}H_{10}N_2O_2}$
(Fig.\ref{model}) was prepared in an analogous fashion using ${\rm
N@C_{60}}$/${\rm C_{60}}$ (around 1/100).

The retention time of ${\rm N@C_{69}H_{10}N_2O_2}$ is slightly
longer than that of ${\rm C_{69}H_{10}N_2O_2}$ in the same way that
${\rm N@C_{60}}$ elutes more slowly than ${\rm
C_{60}}$~\cite{kanai04}. This was demonstrated by measuring the EPR
spin signal of the two halves of the pure product eluting from HPLC.
The first half was found to be spin silent and the second half spin
active.

\subsection{Continuous wave EPR}

\begin{figure}[t] \centerline
{\includegraphics[width=3.2in]{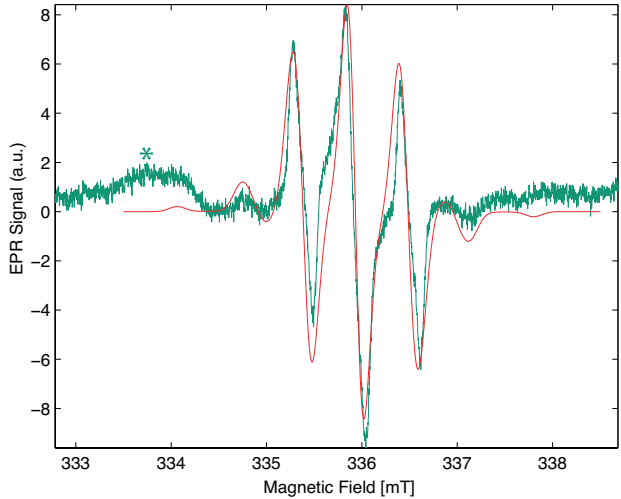}} \caption{X-band CW-EPR spectra of ${\rm
N@C_{69}H_{10}N_2O_2}$ (powder, 2~G modulation, 270~s scan time,
99~scans, room temperature) and their fitting lines (D=17.0~MHz,
E=0.8~MHz, $A_{xx}=A_{yy}=14.2$~MHz, $A_{zz}=18.5$~MHz) (* = impurity).}
\label{cwesr}
\end{figure}

The CW EPR spectrum of ${\rm N@C_{69}H_{10}N_2O_2}$ powder at room
temperature is shown in Fig.~\ref{cwesr}. Satellite peaks on either
side of the three principle lines are indicative of a ZFS. Due to
the low concentration of ${\rm N@C_{69}H_{10}N_2O_2}$ within ${\rm
C_{69}H_{10}N_2O_2}$, the spin exchange interaction between
different molecules can be ignored.

The data is fitted using the spin Hamiltonian:

\begin{equation} \mathcal{H}_0=\omega_e S_z - \omega_I I_z +
\vec{S}\!\cdot\!A \!\cdot\! \vec{I} +
\vec{S}\!\cdot\!\vec{D}\!\cdot\!\vec{S} ,
\end{equation}
where $\omega_e=g\mu_B B_0/\hbar$ and $\omega_I=g_I\mu_N B_0/\hbar$
are the electron and $^{14}$N nuclear Zeeman frequencies, $g$ and
$g_I$ are the electron and nuclear g-factors, $\mu_B$ and $\mu_N$
are the Bohr and nuclear magnetons, $\hbar=h/2\pi$, $h$ is Planck's
constant and $B_0$ is the magnetic field applied along $z$-axis in
the laboratory frame. $A$ is the hyperfine interaction tensor, and
$\vec{D}$ is the ZFS tensor. The ZFS parameters $D$ and $E$
correspond to the eigenvalues $(x, y, z)$ of $D=3z/2$, $E=(y-x)/2$.

The hyperfine coupling in ${\rm N@C_{60}}$ is isotropic and
approximately 15.8~MHz --- about 50$\%$ greater than that of a free
nitrogen atom~\cite{nhyper54}. The fact that the central hyperfine
line (M$_I=0$) in the spectrum shows higher intensity as compared to
the two outer hyperfine lines (M$_I=\pm1$) indicates that the $
^{14}$N hyperfine coupling is anisotropic.

The spectrum is simulated (red line in Fig.~\ref{cwesr}) with ZFS
parameters $D = 17.0$~MHz and $E = 0.8 $~MHz (consistent with
reports on other pyrrolidine-functionalized
fullerenes~\cite{franco06}) and hyperfine terms
$A_{xx}=A_{yy}=14.2$~MHz, $A_{zz}=18.5$~MHz. The functional group
induces a distortion of the fullerene cage, removing the degeneracy
of the encased nitrogen \emph{p}-orbital, resulting in an asymmetric
electron density distribution. This produces the ZFS and an
anisotropy in the hyperfine interaction. The ZFS is about twice that
of the methano-fullerene derivative ${\rm
N@C_{61}(COOC_2H_5)_2}$~\cite{dietel99}, which is consistent with
the higher symmetry of that molecule compared with N@${\rm
C_{69}H_{10}N_2O_2}$.

\subsection{Pulsed EPR}
It has been shown that nuclear spins in the solvent environment
provide a mechanism for electron spin relaxation~\cite{relaxtol}. To
probe the effects of the functional group on spin relaxation, it is
important to use a solvent in which there are no naturally abundant
nuclear spins, such as carbon disulfide.

\begin{figure}[t] \centerline
{\includegraphics[width=3.2in]{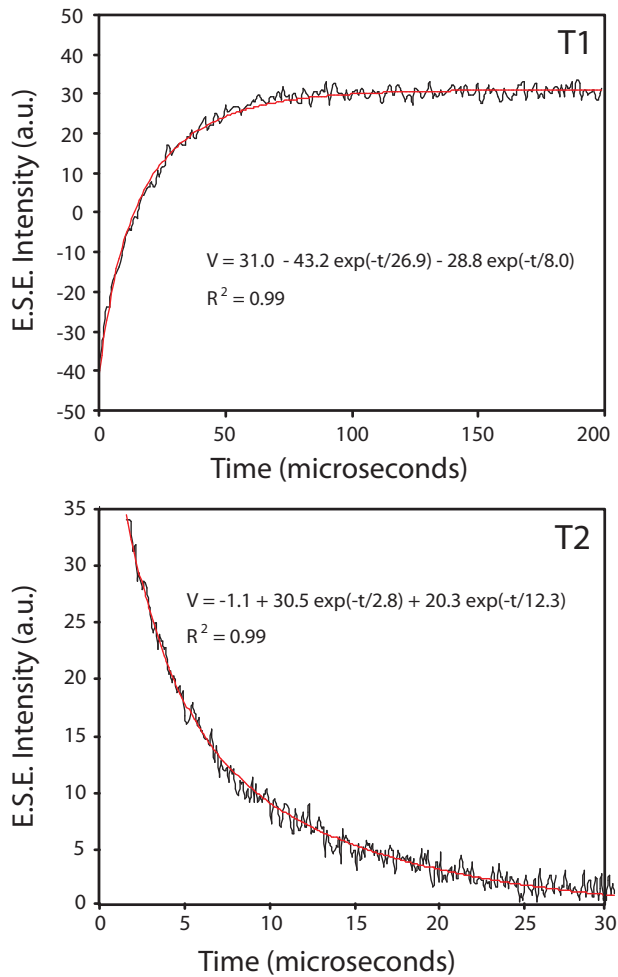}} \caption{Pulsed EPR measurements of ${\rm
N@C_{69}H_{10}N_2O_2}$ in ${\rm CS_2}$ solution at room temperature.
Biexponential fitting lines are derived using equation (2-5) and a
common set of parameters $D_{eff}=17.1\pm0.5$~MHz and
$\tau_c=31.4\pm2.5$~ps.}
\label{pulsedesr}
\end{figure}

Fig.~\ref{pulsedesr} shows relaxation measurements of T$_1$ and
T$_2$ of N@${\rm C_{69}H_{10}N_2O_2}$ in ${\rm CS_2}$ solution at
room temperature, taken using the central (M$_I$=0) hyperfine line.
In both cases, the data were not well fitted by purely
monoexponential decay. Both T$_1$ and T$_2$ have two components
corresponding to the `inner' (M$_S=+1/2:-1/2$) and `outer'
(M$_S=\pm3/2:\pm1/2$) transitions in the $S=3/2$ multiplet. The
relative amplitudes of the `inner' and `outer' contributions to the
total measured echo signal are 2:3. In the case of an $S=3/2$ spin
system, a fluctuating ZFS term will lead to the following relaxation
times~\cite{john06,carrington64}:

\begin{equation} \label{zfst2t2dega} \left(\rm{T_{2,i}}^{-1}\right)_{ZFS}=\frac{4}{5}D_{eff}^2\left[\frac{\tau_c}{1+\omega_e^2\tau_c^2}+\frac{\tau_c}{1+4\omega_e^2\tau_c^2}\right]
\end{equation} \begin{equation} \label{zfst2t2degb}
\left(\rm{T_{2,o}}^{-1}\right)_{ZFS}=\frac{4}{5}D_{eff}^2\left[\tau_c+\frac{\tau_c}{1+\omega_e^2\tau_c^2}\right],
\end{equation} \begin{equation} \label{zfst1t1dega}
\left(\rm{T_{1,i}}^{-1}\right)_{ZFS}=\frac{8}{5}D_{eff}^2\left[\frac{\tau_c}{1+\omega_e^2\tau_c^2}\right]
\end{equation} \begin{equation} \label{zfst2t1degb}
\left(\rm{T_{1,o}}^{-1}\right)_{ZFS}=\frac{8}{5}D_{eff}^2\left[\frac{\tau_c}{1+4\omega_e^2\tau_c^2}\right]
\end{equation} where $D_{eff}^2=D^2+3E^2$, $\tau_c$ is the correlation time of the fluctuations,
and $\omega_e$ is the electron spin transition frequency.

The measured T$_2$ was fitted to a biexponential decay,
$y_0+A_i$exp(-2$\tau$/$T_{2,i}$)+$A_o$exp(-2$\tau$/$T_{2,o}$), which
produced two components with decay rates $12.3\pm0.7$~$\mu$s and
$2.8\pm0.2$~$\mu$s, and relative amplitudes ($A_i/A_o \approx$ 2/3). The
coherence time (T$_2$) of pristine ${\rm N@C_{60}}$ in these
conditions is 80~$\mu$s. Hence, the results suggest an additional
relaxation mechanism affecting the `inner' and `outer' rates
differently, such as that provided by a fluctuating ZFS.

Both T$_1$ and T$_2$ decays shown in Fig.~\ref{pulsedesr} were
fitted using the above equations
(\ref{zfst2t2dega}-\ref{zfst2t1degb}), yielding a molecular rotation
correlation time of $\tau_c=31.4\pm2.5$~ps and a ZFS term $D_{eff} =
17.1\pm0.5$~MHz, matching well with that measured using CW EPR. The
correlation time $\tau_c$ is longer than that of reported fullerene
derivatives and three times that of ${\rm N@C_{60}}$ in
toluene~\cite{dietel99}, showing that the bulky functional group
slows down the tumbling of the fullerene cage. The rotation
correlation time of a somewhat larger nitroxyl derivative of $\rm
C_{60}$ in a benzene-ethanol mixture was reported to be
98~ps~\cite{ivanova04}, measured by EPR of the nitroxyl radical.
This much longer rotational time illustrates the role of a
solvent-solute interaction as an additional factor in molecular
tumbling rates, in which the polar nitroxyl group plays a likely
part. Thus, in molecular dynamics studies of fullerene derivatives,
the use of an endohedral N@C$_{60}$ spin, rather than a nitroxyl
group, as the effective spin label avoids introducing potentially
undesirable interactions with the solvent.

Molecular tumbling in the presence of an anisotropic hyperfine
interaction (HFI) provides an additional relaxation mechanism, which
can be observed by studying the outer hyperfine lines (M$_I=\pm1$).
Based on the hyperfine tensor extracted from CW EPR, the
contribution to phase relaxation from this mechanism is estimated to
be of the order of 100~$\mu$s~\cite{banci91, mcconnel}.

T$_2$ was measured on the low-field (M$_I=+1$) hyperfine line and
found to have a biexponential decay, with components
T$_{2,i}$=$10.2\pm0.7$~$\mu$s and T$_{2,o}$=$2.9\pm0.3$~$\mu$s.
While the effect on the faster T$_{2,o}$ component falls within
experimental error, the effect on T$_{2,i}$ is clear and consistent
with the additional relaxation mechanism of fluctuating hyperfine anisotropy. The effect of ESEEM on this outer
hyperfine line~\cite{john06} is negligible at these short relaxation
times.

\section{Conclusions}

A new pyrrolidine functionalized fullerene derivative, ${\rm
C_{69}H_{10}N_2O_2}$, was synthesized and purified by HPLC. The
structure was confirmed by MALDI mass spectrometry, UV-vis, FTIR,
$^1$H NMR, and $^{13}$C NMR spectroscopies. The magnetic properties
of the analogous species, ${\rm N@C_{69}H_{10}N_2O_2}$, were
measured both by CW and pulsed EPR at room temperature. CW EPR
spectra indicate that the functional group introduces a permanent
ZFS $D=17.0$~MHz and an anisotropy in the hyperfine interaction. The
presence of a ZFS and hyperfine anisotropy was confirmed by the
observation of biexponential decays in both T$_1$ and T$_2$
measurements using pulsed EPR. The nitro functional group was found
to slow down the molecular tumbling in solution, with a correlation
time ${\tau_c}$ of $31.4\pm2.5$~ps at room temperature, three times
that of $\rm N@C_{60}$ in toluene.

\section{Acknowledgements}
This research is part of the QIP IRC www.qipirc.org (GR/S82176/01)
with further support from DSTL. We thank Chris Kay at University
College London for the use of his pulsed EPR spectrometer. We thank
Alexei Tyryshkin at Princeton University for helpful discussions. We
thank EPSRC national mass spectrometry service centre at university
of Wales Swansea. J.Z. is supported by a Clarendon Scholarship,
Overseas Research Student Scholorship and a Graduate Scholarship
from The Queen's College, Oxford. J.J.L.M. is supported by St.
John's College, Oxford. A.A. is supported by the Royal Society.
G.A.D.B. is supported by the EPSRC (GR/S15808/01).

{}

\end{document}